\documentclass[9pt,a4paper,twocolumn]{narms2}

\usepackage{subfigure} % was: subfignarms.sty

\usepackage{psfig}

\usepackage{graphicx}

\usepackage{timesmt}   % use times typeface

\usepackage{chikako}   % <---------------------------------- MOD

\makeatletter

\renewcommand{\section}{\@startsection{section}{1}{0mm}{-
    \baselineskip}{0.15\baselineskip}{\normalfont\normalsize}}

\renewcommand{\subsection}{\@startsection

{subsection}%

{2}%

{0mm}%

{-\baselineskip}%

{0.15\baselineskip}%

{\normalfont\normalsize}}%

\makeatother

\begin{document}

%\maketitle

\title{Aggregation in Charged Nano Suspensions}
\author{
{J.H. Werth \& D.E. Wolf}\\
{\em Fachbereich Physik, Universit\"at Duisburg-Essen, Campus Duisburg, D-47057 Duisburg, Germany}\\[0.3cm]
{S.M. Dammer}\\
{\em Department of Applied Physics, University of Twente, 7500 AE Enschede, The Netherlands}\\[0.3cm]
{H.A. Knudsen}\\
{\em Dept. of Physics, University of Oslo, PB 1048 Blindern, NO - 0316 Oslo, Norway}\\[0.3cm]
{H. Hinrichsen}\\
{\em Fakult\"at f\"ur Physik und Astronomie, Universit\"at W\"urzburg, D-97074 W\"urzburg, Germany}\\[-2cm]
}
\date{}% No date.

\abstract{In order to control aggregation phenomena in suspensions of
nanoparticles, one often charges the particles electrically, e.g. by
triboelectric charging. Stabilization of suspensions against aggregation of
particles is an important issue, which may be realized by monopolar
charging,where particles repel each other. Contrarily, bipolar charging may be
used in coating processes, where smaller particles of one material coat larger
particles of another material. When the two particle fractions are charged
oppositely, aggregation between equally charged particles is hindered whereas
aggregation between oppositely charged particles is preferred, thereby
improving the coating process. We study various aspects of these two
situations by theoretical investigations and computer
simulations.}

%%%%%%%%%%%%%%%%%%%%%%%%%%%%%%%%%%%%%%%%%%%%%%%%%%%%%%%%%%%%%%%%%%%

%  To kick in the changes.
% Make title here for two column format.
\maketitle

\frenchspacing
% no double spaces after colon
% added by <W.Hennings@fz-juelich.de>

% Write some ascii text files called intro.tex, concept.tex, etc.
% TeX and LaTeX will look for the .tex subscript by default.

\section{INTRODUCTION}
Granular materials play a major role in a large number of technological
applications. The size of grains used depends on the application, ranging from
large particles with diameters of several centimeters down to
nanoparticles. However, the physical properties of powders change drastically
when reaching the size of nanoparticles. While in ordinary dry granular
materials contacts between the particles cannot sustain tensile loads, in the
case of ultra-fine powders van der Waals forces lead to aggregation which is
essentially irreversible. In contrast to volume forces such as gravity, which
decrease with decreasing particle radius $a$ like $a^{3}$, van der Waals
forces are proportional to $a$ and thus dominate all other forces in the case
of sufficiently small particles. On the other hand, van der Waals forces are
only short ranged, diverging like $h^{-2}$ on very short
distances and decaying like $h^{-7}$ with increasing distance $h$ between the surface
of particles \shortcite{van_de_ven89}. Thus, the van der Waals force
can be approximated as an irreversible sticking force, which has no
influence on particles separated from each other, but causes them to stick together irreversibly, as soon as they
touch each other.

A promising strategy to reduce agglomeration of particles is to charge them electrically. If one manages to charge all particles with the same
polarity, the mutual Coulomb repulsion delays or even suppresses the
aggregation process. It has been shown that such a charge distribution can be
established for particles suspended in a non-polar liquid
\shortcite{werth03,linsenbuehler05}. In the first part of this paper
we investigate to what extent charges can be used to enhance the stability
of a suspension against agglomeration.
We present some results obtained from a rate equation investigation, recently
published in \shortcite{stephan_prl}.

In the second part of the paper, we present an application which makes use of
Coulomb forces: the coating of charged particles with sizes on the micrometer
scale by oppositely charged nanoparticles. As shown in \shortcite{markus_partec},
such a coating may enhance the flowability of a powder dramatically, since the
nanoparticles on the particle surfaces act as spacers between the micrometer
sized particles, thus reducing their mutual van der Waals forces. The
aggregation process of oppositely charged particles has been studied in
detail~\shortcite{ladungsfindung}.

\section{STABILIZATION AGAINST AGGLOMERATION}
Let us consider a very fine suspended powder. The size of the particles shall be a few
$\mu$m or below. We assume the particles to be all of the same insulating
material, each carrying an
electric charge of the same sign. The charged particles are subjected to Coulomb and van der Waals forces as
well as Brownian motion and Stokes friction. As already pointed out in the introduction, van der
Waals forces will only dominate over Coulomb forces for particles very close
to each other. Thus we assume that the van der Waals force has no effect except when particles collide and stick together irreversibly. The motion of the particles is then composed of a stochastic part due to Brownian
motion and a deterministic part determined by the balance of repelling Coulomb
forces and Stokes friction~\shortcite{buch}. It may happen that particles (or
agglomerates of particles) come into contact due to Brownian motion, thus
forming a new agglomerate (Brownian coagulation). In the following we will investigate this
aggregation process in further detail in terms of a rate equation approach.

We assume that initially the
suspension consists of primary particles or primary aggregates, all of the same mass~$m^*$, the
same effective radius~$a^*$, and carrying
the same charge $q^*$. As particles collide they form clusters with increasing
mass and charge. Since for the chosen initial conditions mass and charge of the clusters are proportional to
each other, it is sufficient to describe them by a single index
$i$ according to
\begin{equation}
\label{coagulation_eq}
\frac{dn_i(t)}{dt}=\frac{1}{2}\sum
_{j+k=i}R_{jk}n_j(t)n_k(t)-n_i(t)\sum
_{j=1}^{\infty}R_{ij}n_j(t).
\end{equation}
Here $n_i(t)$ denotes the number density of clusters with
mass
$m_i{=}im^*$ and radius $a_i{=}i^\alpha a^*$ at time
$t$,
each of them carrying the charge $q_i{=}iq^*$,
where
$1/\alpha$ denotes the fractal dimension of the aggregates
(e.g.
$\alpha{=}1/3$ for spherical particles). As initial
condition we choose $n_1(t{=}0){=}1,\ n_i(t{=}0){=}0\mbox{\ for\ }i{>}1\,$.
The matrix $R_{ij}$ in the rate equation is called coagulation
kernel
and describes at which rate two clusters with indices $i$
and
$j$ merge into a single one. For
Brown\-ian coagulation of charged particles one has
\begin{eqnarray}
R_{ij}&=&(i^{\alpha}+j^{\alpha})(i^{-\alpha}+j^{-\alpha})\,\frac{\kappa_{ij}}{\exp(\kappa_{ij})-1}\,,
\label{rate_prototype_1}\\
\kappa_{ij}&=&\frac{k^2ij}{(i^\alpha+j^\alpha)}\label{kappa}\
,
\end{eqnarray}
where $k^2{=}{q^*}^2/(4\pi\epsilon _0\epsilon _{\rm r} a^*k_{\rm B}T)$
and
$2k_{\rm B}T/3\eta{=}1$ \shortcite{stephan_prl}. Here $\kappa_{ij}$ is proportional to the Coulomb
energy
of two clusters being in contact, divided by the thermal energy.
% neu:
This Coulomb energy is necessary to bring two particles into contact from infinite separation. If the suspension contains countercharges so that a double layer forms, the Coulomb repulsion at short distances is combined with attractive forces at intermediate distances. In this case the activation energy for a particle collision can be enhanced, i.e. the collision rate can be reduced compared to Eq. (\ref{rate_prototype_1}).

\begin{figure}
\centerline{\includegraphics[width=60mm]{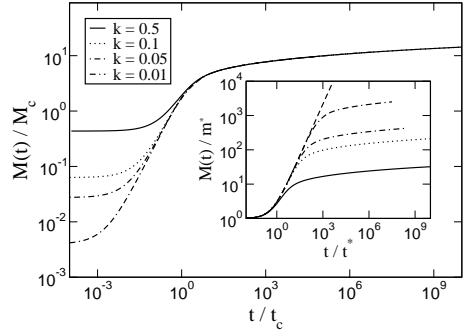}}
\caption{\small
\label{crossover}
The average mass $M(t)$ scaled with $M_{\rm c}$ vs. time $t$ scaled with
$t_c$ for different values of $k$. The data collapse shows that in all cases the
crossover to slow cluster growth happens at $t{=}t_c$ and $M{=}M_c$.
The inset shows the original data, and the dashed line
is the behavior for an uncharged system ($k{=}0$).}
\end{figure}

If the system is initially unstable,
i.e. $k^2{\ll}1$,
it behaves essentially as if it was uncharged and aggregation occurs
frequently. In this regime, the average aggregate mass increases linearly in
time~\shortcite{Ernst}. However, as soon as particles with $\kappa_{ij}{\approx} 1$ become
important, further aggregation is suppressed
exponentially. Asymptotically, in this regime the average aggregate mass
increases sub-logarithmically slow in time~\shortcite{stephan_prl}. The crossover
between these two regimes happens at a characteristic time $t_c$ and
mass $M_c$ given by
\begin{equation}
t_c \approx
k^{-\frac{2}{2-\alpha}}t^*\ \ ,\ \ M_c\approx k^{-\frac{2}{2-\alpha}}m^*\ ,
\end{equation}
where $t^*$ is the appropriate time unit, given by
$t^*=3\eta/\tilde{n}2k_{\rm B}T$. $\tilde{n}$ is the initial number
density of primary particles. Fig.~\ref{crossover} shows the temporal evolution of
the average aggregate mass, obtained from a numerical solution of the rate
equations~(\ref{coagulation_eq}).
As an example, for an experimental situations
with values like $\eta{=}10^{-4}{\rm Pa\,s}$, $T{=}300{\rm K}$, $a^*{=}1{\rm
  \mu m}$ and $\tilde{n}{\approx}2{\cdot}10^5{\rm mm}^{-3}$ one finds
$t^*{\approx}150{\rm s}$. Assuming $\epsilon _{\rm r}{\approx}1$,
charging with a single elementary charge on each primary particle
corresponds to a value $k{\approx}0.2$. For these parameters one finds
with the dimensionless results of Fig.~\ref{crossover} that the crossover to
slow aggregation occurs within hours.

For times larger than $t_c$ the suspension is usually called stable. However,
note that aggregation still proceeds slowly. A consequence of this is that the
mass distribution evolves towards a {\em universal} scaling form and that the
relative width $\sigma _r$ of the distribution falls off to a universal value $\sigma
_r^{\infty}{\approx}0.2017$ that is much smaller than in the uncharged
case (with $\sigma _r{\approx} 1$)~\shortcite{stephan_prl}. Hence, for an initially narrow mass
distribution $\sigma _r$ first grows similarly to the uncharged situation until
time $t_c$, when it starts decreasing again (induced by the exponential
suppression of further aggregation) which is denoted as {\em
  self-focusing}. With the values of the example discussed above, this
self-focusing effect occurs within an experimentally accessible time.

%hint: for citations according to Balkema style, use "\shortcite{key}"
%      instead of "\shortcite{key}

\section{COATING OF PARTICLES WITH NANOPARTICLES}
In this section we present an application which makes use of
charged granular particles in suspension, namely the coating of powder
particles with nanoparticles. As shown by experiments \shortcite{markus_partec},
such a coating of powder particles, even by a small amount of nanoparticles,
leads to a strong increase in flowability and fluidizability of
powders.

A detailed description of the coating process can be found in
\shortcite{werth03}. Therefore, we want to give only a brief description. During
the process, both the powder particles and the nanoparticles are charged
triboelectrically by stirring the suspension with a high speed stirring
device. By choosing different, electrically insulating materials for the
powder particles and the nanoparticles it is possible to charge both fractions
oppositely in a single process. After the charging process is stopped, the
suspension is left untouched and the particles may agglomerate with each
other. Due to Coulomb forces, agglomeration of oppositely charged particles
(i.e. powder particles and nanoparticles) is preferred, while agglomeration of
equally charged particles is suppressed. Eventually, a dry powder of coated
particles can be achieved by evaporation of the liquid
nitrogen. 

An interesting question is, to what extent the charges influence the
deposition of the nanoparticles on the larger powder grains. To this end, we
expect the nanoparticles to usually carry only one single elementary charge,
while the larger powder particles may carry several elementary charges. Since
the used materials are electric insulators, the elementary charges are
immobilized at distinct positions on the particle surfaces. If a nanoparticle
now approaches the surface of a larger particle, two situations may arise: it
may touch the particle exactly at the position of a charge, thereby
compensating this charge, or it may just touch on a random position on the
surface, leaving a dipole moment interacting with more charged nanoparticles
in the suspension. The second case may occur due to Brownian motion of the
particles.

To answer the question, which situation is more likely and to what extent
surface charges of powder particles are compensated by charged nanoparticles,
we study a simplified model: point-like particles, each carrying one negative
elementary charge, are continously inserted far away from one infinitely large
particle, represented by a flat wall and carrying one or several positive
elementary charges. The point-like particles are subjected to Coulomb forces by
interaction with the charges on the wall, to Brownian motion and to Stokes
friction by the surrounding fluid. Coulomb interaction of equally
charged particles among each other is not regarded. As soon as a particle touches the wall, it
is removed from the system. Our aim is to compute the probability distribution
$\rho(\vec{r})$ for the particles to touch the wall at position
$\vec{r}$.

The answer to this problem is given in detail in
\shortcite{ladungsfindung}. Therefore, here we only motivate the according
equations and present some results. The probability distribution
$P(\vec{r},t)$ to find a particle at point $\vec{r}$ and time
$t$
can be described by a Fokker-Planck (FP) equation. It has the
form
\begin{equation}
\frac{\partial}{\partial t} P(\vec{r},t) = -\vec{\nabla}\cdot
\vec{j}(\vec{r},t)\,,
\end{equation}
where
\begin{equation}
\vec{j}(\vec{r},t) = -D\vec{\nabla} P (\vec{r},t) + \vec{v}(\vec{r}) P
(\vec{r},t)
\end{equation}
is the probability current, $D$ the diffusion constant,
and
\begin{equation}
\vec{v}(\vec{r}) = -\frac{q^2 \vec{r}}{24 \pi^2 \varepsilon \varepsilon_0 \eta
a r^3}
=
-\frac{Q}{r^2} \,
\frac{\vec{r}}{r}
\end{equation}
is the particle velocity in the overdamped
limit.

Measuring space in units of $Q/D$ and time in units of $Q^2/D^3$
we obtain the parameter-free dimensionless
equation
\begin{equation}
\label{DimlessFP}
\frac{\partial}{\partial t} P
=
\nabla^2 P
-
\vec{\nabla} P\cdot
\vec{u}-
P \, (\vec{\nabla} \cdot
\vec{u})
\end{equation}
where
$\vec{u}=-\vec{r}/r^3$.

We are only interested in the distribution $\rho(\vec{r})$ of incoming
particles at the
wall.
This allows us to use a time independent FP equation. We arrive at the
stationary FP
equation
\begin{equation}
\nabla^2P - \vec{\nabla}P\cdot\vec{u}-P(\vec{\nabla}\cdot \vec{u}) = 0
.
\label{stationaryfpequation}
\end{equation}

Equation (\ref{stationaryfpequation}) can be solved analytically in the case
of one single charge fixed at the wall. The analytical solution as well as an
in-depth derivation of (\ref{stationaryfpequation}) and a numerical treatment
of the according Langevin equation can be found in
\shortcite{ladungsfindung}.

First we want to concentrate on the analytically solvable case of one single
surface charge at the wall. The density distribution $\rho(r)$ of particles
reaching the wall in a distance $r$ to the fixed charge is given
by
\begin{equation}
\rho(r)=1+\frac{1}{2r}+\frac{\pi}{2}\delta(r)
,
\label{solution_singlecharge}
\end{equation}
where the density of nanoparticles far away from the wall is set to
unity. There are three terms contributing to the density distribution:
Firstly, there is a certain fraction of point particles which exactly reaches
the surface charge, represented by the term $\frac{\pi}{2}\delta(r)$. Secondly,
there is a concentration of particles around the fixed charge, decaying like
$1/2r$ with growing distance $r$ to the fixed charge. Finally, there is a
constant background of incoming particles according to the density of
particles far away from the wall. Solution (\ref{solution_singlecharge})
provides no intrinsic length
scale.

\begin{figure}
\centerline{\includegraphics[width=60mm]{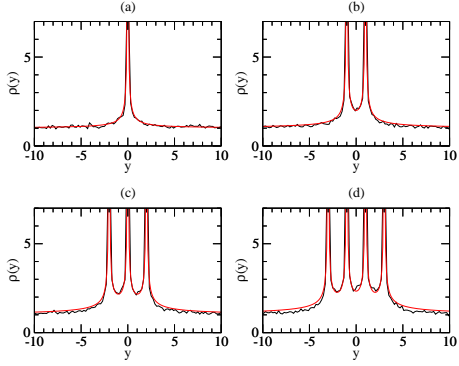}}
\caption{\small
\label{lokale_auswertungen}
Density profile of particles hitting the wall. One (a), two (b), three (c) and
four (d) charges are fixed on the wall in a line. The pictures show density
profiles of incoming particles along this line. The bold curves present
numerical results, while theoretical predictions are plotted in light
grey.
}
\end{figure}

Moving from one single fixed charge to several charges distributed at the
wall, there is no analytical solution of equation (\ref{stationaryfpequation})
until now. Especially, the sum of solutions of type
(\ref{solution_singlecharge}) for several charges at different positions is
not a solution of the multi-charge Fokker-Planck equation. However, as we will
see, at least for only a few charges, adding up the single charge solutions
can be a first approximation for the real
problem.

Since analytical solutions are not tangible, the direct numerical simulation
of a multi-charge Langevin equation can be used to obtain results. The
Langevin equation describing the movement of a nanoparticle is given
by
\begin{equation}
\frac{\partial}{\partial t}\vec{r}=\frac{\vec{F}_{\rm C}(\vec{r})}{6\pi\eta
a}+\vec{\xi}(t)
\label{langevin_equation}
\end{equation}
where $\vec{r}$ is the position of the particle, $\vec{F}_{\rm C}(\vec{r})$ is
the Coulomb force acting on it, $\eta$ is the dynamic viscosity of the fluid,
$a$ the particle radius and $\vec{\xi}(t)$ is a white Gaussian noise corresponding to Brownian motion of the particle.

Fig.~\ref{lokale_auswertungen} shows densities of particle influx at the wall
obtained numerically. In the simulations corresponding to Figures
\ref{lokale_auswertungen} (a)-(d), up to four charges are arranged in a
line, separated by a distance of 2$Q/D$. The bold curves show
the density of particles touching the wall along this line. The light curves are obtained by adding up the $1/2r$-shoulders
from the single charge solution (\ref{solution_singlecharge}) for all
charges. If we denote the position along the line of charges by $y$, the light
curves are given
by
\begin{equation}
\rho(y)=1+\sum\limits_{i=1}^n\frac{1}{2|y-y_i|},
\end{equation}
where $n$ is the number of charges fixed at the wall. This approximation
agrees nicely with simulations, although in the case of three and four charges
the numerical data lie slightly below the light
curves.

\section{CONCLUSION}
We discussed suspensions of monopolarly and bipolarly charged particles. In
the former case, accumulation of charges leads to a crossover from fast to
slow aggregation, followed by a self-focusing of the mass distribution towards
a universal scaling form. In the latter case, we showed that bipolar charging may enhance the coating of powder particles with nanoparticles.\\[0.5cm]

{\noindent\bf Acknowledgements}\\
We thank Z. Farkas, M. Linsenb\"uhler and K.-E. Wirth for fruitful discussions. This work was supported by the German Science Foundation (DFG) within the research program ''Verhalten Granularer Medien'', project Hi/744.\\

%\newpage

\bibliographystyle{chikako}    % <----------------------------------MOD
\bibliography{paper} % These is my BiBTeX reference list.
\end{document}